\documentclass[conference]{IEEEtran}
\IEEEoverridecommandlockouts
\usepackage{cite}
\usepackage{amsmath,amssymb,amsfonts}
\usepackage{graphicx}
\usepackage{textcomp}
\usepackage{xcolor}
\usepackage{subfigure}
\usepackage{enumerate}
\usepackage{algorithmic}
\usepackage[ruled,vlined,linesnumbered]{algorithm2e}
\SetKwInput{KwInput}{Input}
\SetKwInput{KwOutput}{Output}

\def\BibTeX{{\rm B\kern-.05em{\sc i\kern-.025em b}\kern-.08em
    T\kern-.1667em\lower.7ex\hbox{E}\kern-.125emX}}
\begin{document}

\title{Enhancing Cloud Task Scheduling Using a Hybrid Particle Swarm and Grey Wolf Optimization Approach}

\author{
\IEEEauthorblockN{Raveena Prasad}
\IEEEauthorblockA{
\textit{Computer Science and Engineering} \\
\textit{Shiv Nadar Institute of Eminence} \\
\textit{Delhi-NCR, India} \\
rp131@snu.edu.in}
\and
\IEEEauthorblockN{Aarush Roy}
\IEEEauthorblockA{
\textit{Computer Science and Engineering} \\
\textit{Shiv Nadar Institute of Eminence} \\
\textit{Delhi-NCR, India} \\
ar833@snu.edu.in}
\and
\IEEEauthorblockN{Suchi Kumari}
\IEEEauthorblockA{
\textit{Computer Science and Engineering} \\
\textit{Shiv Nadar Institute of Eminence} \\
\textit{Delhi-NCR, India} \\
suchi.singh24@gmail.com}
}

\maketitle

\begin{abstract}
Assigning tasks efficiently in cloud computing is a challenging problem and is considered an NP-hard problem. Many researchers have used metaheuristic algorithms to solve it, but these often struggle to handle dynamic workloads and explore all possible options effectively. Therefore, this paper presents a new hybrid method that combines two popular algorithms, Grey Wolf Optimizer (GWO) and Particle Swarm Optimization (PSO). GWO offers strong global search capabilities (exploration), while PSO enhances local refinement (exploitation). The hybrid approach, called HybridPSOGWO, is compared with other existing methods like MPSOSA, RL-GWO, CCGP, and HybridPSOMinMin, using key performance indicators such as makespan, throughput, and load balancing. We tested our approach using both a simulation tool (CloudSim Plus) and real-world data. The results show that HybridPSOGWO outperforms other methods, with up to 15\% improvement in makespan and 10\% better throughput, while also distributing tasks more evenly across virtual machines. Our implementation achieves consistent convergence within a few iterations, highlighting its potential for efficient and adaptive cloud scheduling.
\end{abstract}

\begin{IEEEkeywords}
Cloud computing, Task scheduling, Grey Wolf Optimizer, Particle Swarm Optimization, Metaheuristics, Load balancing
\end{IEEEkeywords}

\section{Introduction}
Cloud computing has revolutionized the IT landscape by enabling flexible, scalable, and on-demand resource provisioning. As cloud infrastructures grow, efficient task scheduling becomes increasingly important to ensure high system performance, balanced workload distribution, and optimal resource utilization. Researchers have proposed multiple approaches to optimize some objective functions. Given the complexity and dynamic nature of cloud environments, it cannot be solved using traditional mathematical approaches such as the Gradient Descent Approach, Quasi-Newton Methods, and some other optimization schemes \cite{bellman1963mathematical}. We need some metaheuristic algorithms to tackle multimodal, non-continuous, and non-differentiable problems. Traditional heuristics often fall short due to their limited ability to explore large search spaces or adapt to evolving workloads. 

Metaheuristic algorithms can be broadly classified into two main types: single solution-based methods, such as Simulated Annealing (SA) \cite{rutenbar1989simulated}, and population-based methods \cite{beheshti2013review}. Single solution-based algorithms work with only one solution at a time, while population-based methods are inspired by natural processes and work with a group of solutions. These algorithms generate new populations by modifying existing ones over several iterations until a stopping condition is met.
Population-based metaheuristics can be further categorized based on their underlying principles. For example, Evolutionary Algorithms (EAs) \cite{bartz2014evolutionary}, Gravitational Search Algorithms (GSA) \cite{rashedi2009gsa}, and algorithms based on Swarm Intelligence \cite{chakraborty2017swarm}.
Despite the diversity of these algorithms, they all follow a common structure involving two main phases: exploration (or diversification) and exploitation (or intensification). During the exploration phase, the algorithm searches different regions of the solution space to discover promising areas. In the exploitation phase, the focus shifts to refining and improving the best solutions found so far. Maintaining a proper balance between these two phases is crucial, as the algorithm may get stuck in local optima and fail to provide the global best solution.

Some metaheuristic algorithms, such as Particle Swarm Optimization (PSO), Grey Wolf Optimizer (GWO), Genetic Algorithms (GA), and Ant Colony Optimization (ACO), offer flexible search mechanisms suitable for complex optimization problems. PSO is a widely used algorithm inspired by social behaviors in swarms, which excels in convergence speed but is prone to premature convergence in multimodal landscapes. GWO, modeled on the leadership hierarchy and hunting strategies of grey wolves, has shown strong exploration capabilities but may lack fine-tuning accuracy in the exploitation phase. GA mimics natural selection but can sometimes get trapped in local optima, especially with very complex problems. ACO is inspired by how ants find the shortest path in search of food. However, it doesn't explore new solutions very well and is often slower to find the best result.

To overcome the limitations of using individual metaheuristic algorithms, hybrid approaches have become increasingly popular. These approaches combine algorithms that are strong in exploration with those that perform well in exploitation. Some studies have developed hybrid methods by carefully tuning both global and local parameters. For example, combinations like PSO with Simulated Annealing and Reinforcement Learning with GWO have been explored to balance exploration and exploitation. However, these methods still struggle with adaptability and efficient convergence. To address these issues, this paper introduces a new hybrid metaheuristic called HybridPSOGWO, which combines the strengths of both GWO and PSO. GWO provides strong global search ability, while PSO improves local refinement. By integrating these two, the proposed method aims to achieve better performance in task assignment for cloud computing. The main contributions of our work include:

\begin{itemize}
    \item Provide a novel integration mechanism that synergistically combines PSO's velocity-based updates with GWO's hierarchical leadership structure.
    \item Consider a VM-awareness component that considers capacity constraints during task assignment.
    \item Implementation in both CloudSim Plus simulation environment and with real-world Google Borg traces.
    \item Comprehensive evaluation against state-of-the-art scheduling algorithms using multiple performance metrics.
\end{itemize}

The manuscript is organized as follows. Section \ref{Sec2} reviews relevant metaheuristic algorithms and existing hybrid models. Section \ref{Sec3} describes the architecture and details of the proposed HybridPSOGWO model. The results and in-depth analysis are presented in Section \ref{Sec4}.  Section \ref{Sec5} discusses the effectiveness, practical implications, and challenges of the proposed approach. Finally, Section \ref{Sec6} concludes the paper and outlines the future direction.

\section{Related Work} \label{Sec2}
Researchers have provided multiple multi-objective heuristic and metaheuristic-based task scheduling approaches. The main goal of such approaches is to maintain load balance, reduce the makespan time, enhance throughput, and efficiently utilize cloud resources. 
\subsection{Metaheuristic Approaches for Task Scheduling}
Task scheduling in cloud computing has been addressed using various metaheuristic algorithms. Lipsa \textit{et al.} \cite{lipsa2023task} reduced the task waiting time by assigning priority to each task and extracted the highest priority task using Fibonacci Heap. Devi \textit{et al.} \cite{devi2024systematic} review task scheduling techniques, highlighting the strengths of hybrid approaches in improving scheduling efficiency. Huang \textit{et al.} \cite{huang2024task} propose a GWO-based method with a novel encoding mechanism that improves task allocation and reduces execution time. Mirjalili \textit{et al.} \cite{mirjalili2014grey} models the social hierarchy and hunting behavior of grey wolf packs. In cloud computing, GWO variants have shown promising results by maintaining better population diversity. Gupta \textit{et al.} \cite{gupta2017improved} proposed an improved GWO for workflow scheduling in cloud environments, demonstrating better convergence than standard GWO.

Kennedy \textit{et al.} \cite{kennedy1995particle} proposed particle swarm optimization (PSO) that has been widely applied to cloud scheduling due to its simplicity and fast convergence. However, standard PSO implementations often struggle with local optima traps in complex scheduling scenarios. Huang \textit{et al. }\cite{huang2020task} considered variants of PSO and compared the makespan time with other existing algorithms, such as GSA, artificial bee colony algorithm, and dragonfly algorithm, respectively. Wang \textit{et al.}\cite{wang2021effective} integrate PSO with idle time slot-aware rules to optimize makespan and reduce resource wastage. 

\subsection{Hybrid Metaheuristic Approaches}
Hybrid metaheuristics combine multiple algorithms to overcome the limitations of individual approaches.  Xiao \textit{et al.} \cite{xiao2019cooperative} introduce a cooperative coevolution hyper-heuristic framework (CCGP) that combines multiple heuristics to improve scheduling efficiency. Reinforcement learning has also been integrated with metaheuristics. Tong et al. \cite{tong2020scheduling} proposed RL-GWO, which uses reinforcement learning principles to guide GWO's search process, showing improved adaptability in dynamic environments. They scheduled the tasks dynamically using deep Q-learning task scheduling (DQTS), which combines the advantages of the learning algorithm and a deep neural network. Pirozmand \textit{et al.} \cite{pirozmand2021multi} discuss a multi-objective hybrid genetic algorithm to optimize both makespan and resource utilization. Ghahari \textit{et al.} \cite{ghahari2025efficient} proposed a joint approach for task offloading and dynamic resource scaling for handling the dynamic traffic using a deep reinforcement learning approach. To balance the load among all the distributed servers, Kathole \textit{et al.} \cite{kathole2025novel} propose a multiobjective-based Random Opposition Coati Optimization Algorithm for resource prediction and applied a federated learning based secure resource distribution using a blockchain framework. 

Pirozmand \textit{et al.} \cite{pirozmand2023improved} identified an ideal timetable for task scheduling to reduce the makespan time as well as system efficiency. They proposed a Multi-Adaptive Learning for Particle Swarm Optimization (MALPSO) to bifurcate the particle into ordinary particles and locally best particles, to reduce the variety of population and time to reach the optima.  Khan \textit{et al.} \cite{khan2022task} proposed a hybrid optimization algorithm to schedule the jobs effectively with the least amount of waiting time. Mansouri et al. \cite{mansouri2019hybrid} proposed a hybrid PSO with MinMin heuristic to improve local search capabilities. Similarly, Agarwal and Srivastava \cite{agarwal2018improved} combined PSO with simulated annealing (MPSOSA) for task scheduling, showing improved exploration and exploitation balance. Fu \textit{et al. }\cite{fu2023task} proposed a hybrid particle swarm algorithm and genetic algorithm to reduce the completion time of tasks with a higher convergence accuracy. Ramesh \textit{et al.} \cite{Ramesh2025HGWO} combines Grey Wolf Optimization (GWO) with Simulated Annealing (SA) to enhance workflow scheduling in cloud environments. GWO provides efficient global search, while SA improves local exploitation and avoids local optima. This hybrid approach achieves better makespan, cost, and load balancing compared to PSO, GA, and GSA. Despite these advancements, existing hybrid approaches still face challenges in achieving the right balance between exploration and exploitation, handling VM capacity constraints, and adapting to different workload characteristics.

\section{Proposed Approach} \label{Sec3}
This section presents the architectural diagram of the proposed Hybrid PSOGWO system. The task scheduling problem is formally defined, followed by a detailed description of the algorithmic steps. The integration of adaptive mechanisms, which effectively balances exploration and exploitation, while preserving population diversity and ensuring load balancing, is also discussed. The particle positions in PSO are updated using specific policies designed to enhance performance.
\subsection{System Architecture}
\begin{figure*}[htbp]
\centering
\includegraphics[width=\textwidth, height = 3.5 in]{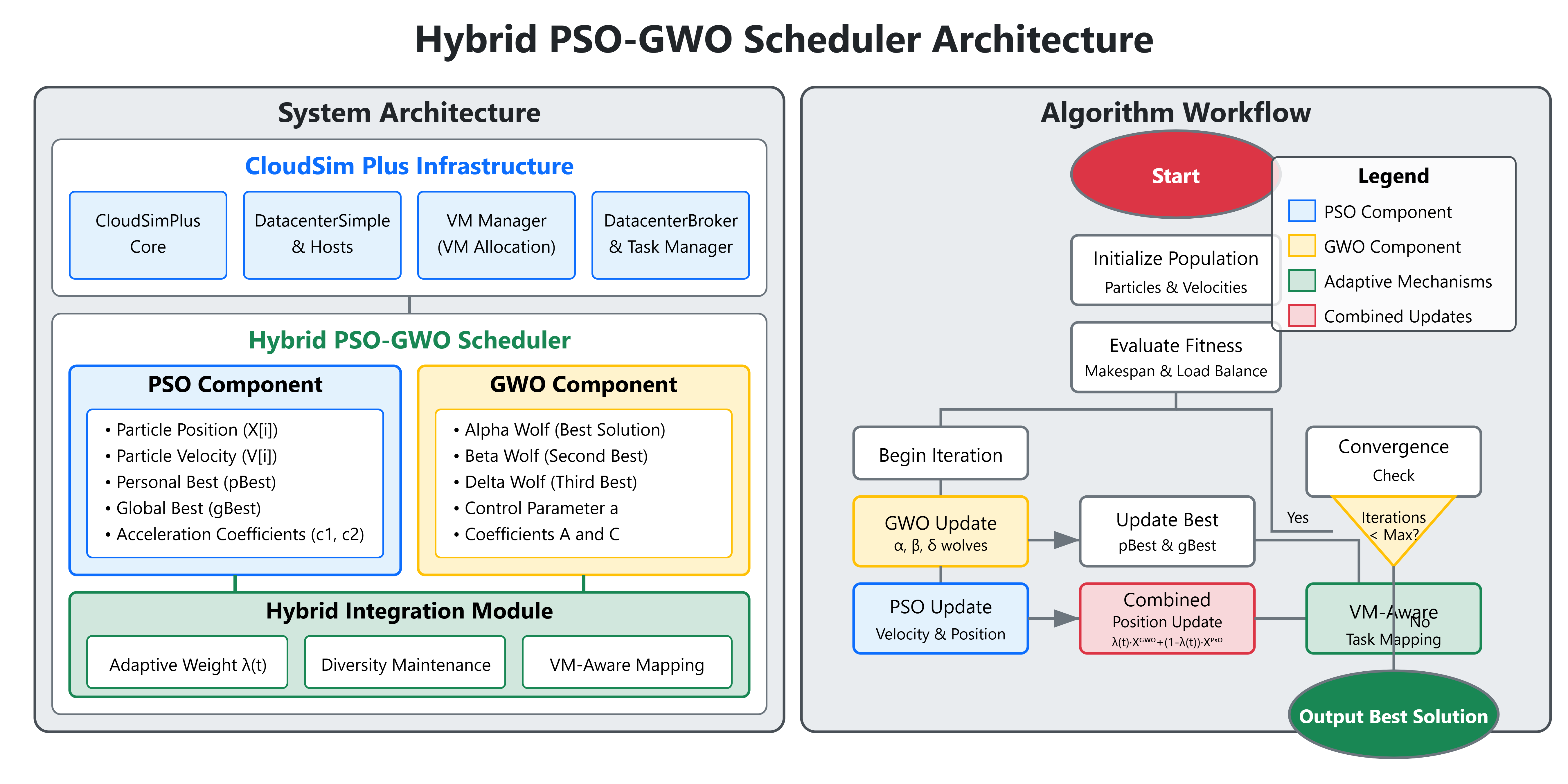}
\caption{Architecture of the proposed HybridPSOGWO approach for cloud task scheduling, showing the integration of PSO and GWO components with VM-aware task mapping.}
\label{fig:architecture}
\end{figure*}

The architecture of our proposed HybridPSOGWO approach is illustrated in Fig. \ref{fig:architecture}. The system comprises the following key components.

\begin{itemize}
    \item \textbf{Task Manager:} The Task Manager is responsible for receiving and managing incoming tasks along with their requirements. It serves as the entry point for tasks submitted to the cloud environment and maintains a queue of tasks awaiting scheduling.
    \item \textbf{VM Manager:} The VM Manager maintains detailed information about available virtual machines (VMs) and their capacities. This component tracks resource availability, processing capabilities, and current load status of each VM in the cloud environment.
    \item \textbf{HybridPSOGWO Scheduler:} This is the core component of the architecture that integrates PSO velocity updates with GWO's hierarchical leadership structure. The scheduler processes the task queue and determines optimal task-to-VM assignments based on the hybrid metaheuristic approach.
    \item \textbf{Position Encoder/Decoder:} This component handles the mapping between continuous position values in the algorithm's search space and discrete VM assignments. It transforms the continuous values produced by the metaheuristic into practical VM assignments for task execution.
    \item \textbf{VM-Aware Task Mapper:} The VM-Aware Task Mapper assigns tasks to VMs while considering capacity constraints. It ensures that tasks are distributed to appropriate VMs without exceeding resource limitations, improving load balancing and preventing resource overutilization.
    \item \textbf{Performance Monitor:} The Performance Monitor evaluates and tracks scheduler performance using defined metrics such as makespan, throughput, and load balance. This component provides feedback to the scheduler to adapt and improve task assignments over time.
    \item \textbf{Adaptive Weight Module:} This specialized component dynamically tunes the $\alpha(t)$ parameter from 0.9 to 0.4 over iterations. The adaptive weight controls the balance between PSO and GWO components throughout the optimization process, shifting from exploration-focused to exploitation-focused behavior. 
    \item \textbf{Diversity Monitor:} The Diversity Monitor computes the average inter-particle distance within the population and injects Gaussian mutation when diversity falls below a defined threshold. This mechanism prevents premature convergence and ensures continued exploration of the solution space.
    \item \textbf{VM-Aware Discretizer:} This component maps continuous score values produced by the algorithm into valid VM indices while respecting capacity constraints. It ensures that the final task assignments are practical and implementable in the actual cloud environment.

\end{itemize}

The scheduler leverages PSO's cognitive and social learning capabilities alongside GWO's alpha, beta, and delta leadership hierarchy to guide the search for optimal task assignments.

\subsection{Problem Formulation}
We define the cloud task scheduling problem as follows. Suppose we have a set of tasks $T = \{t_1, t_2, \ldots, t_n\}$ and a set of virtual machines $VM = \{vm_1, vm_2, \ldots, vm_m\}$. The goal is to assign all $n$ tasks to the $m$ VMs in a way that optimizes certain performance objectives such as reducing completion time, improving resource usage, and balancing the load.

\begin{itemize}
    \item \textbf{Makespan:} It refers to the total time required to complete all the tasks in an application. To execute a task, it must be mapped to a virtual machine (VM). To evaluate the makespan time, we need to find the maximum completion time across all VMs. An objective function is formulated to minimize the makespan time in Eq. \eqref{minmake}.
\begin{equation}
\begin{aligned}
\min Makespan = \quad &  \max_{{i \in \{1,2,\ldots,n\}}, {j \in \{1,2,\ldots,m\}}} \{CT_{i,j}\} \\
\textrm{subject to} \quad & CT_{i,j} = exit(t_{i,j}) - entry(t_{i,j}) \\
\quad &CT_{i,j} \neq 0, \textrm{     } \textrm{if }   t_i\textrm{ 
 is mapped to  } vm_j  \\
\quad &CT_{i,j} = 0, \textrm{ Otherwise }\label{minmake}
\end{aligned}
\end{equation}
where $CT_{i,j}$ is the completion time of task $t_i$ on a virtual machine $vm_j$. To evaluate the completion time of a task, we need to know the entry time, $entry(t_{i,j})$, and exit time, $exit(t_{i,j})$, of task $t_i$ on $vm_j$. The aim of the proposed objective function is to minimize the makespan time.

\item \textbf{Throughput:} Throughput refers to the number of tasks completed per unit of time. Eq. \eqref{minmake} is focused on minimizing the makespan, which helps in evaluating the throughput of the algorithm. In Eq. \eqref{maxthrouhput}, we define an objective function aimed at maximizing throughput when all $n$ tasks are executed across the $m$ virtual machines.

\begin{equation}
\begin{aligned}
\max Throuhput = \quad &  \frac{n}{Makespan} \\
\textrm{subject to} \quad & Makespan \neq 0\label{maxthrouhput}
\end{aligned}
\end{equation}

\item \textbf{Load balance:} To ensure that all virtual machines (VMs) are used efficiently, tasks should be evenly distributed across the $m$ VMs. To ensure this, we formulated an objective function to minimize the coefficient of variation (CV) as shown in Eq. \eqref{eqCV}. 

\begin{equation}
\min CV = \frac{\sigma(Load)}{\mu(Load)} \label{eqCV} 
\end{equation}

where $\sigma(Load)$ represents the standard deviation of VM loads and $\mu(Load)$ is the average load. A lower CV value indicates a more balanced distribution of tasks among the VMs, which leads to better resource utilization.

\item \textbf{Solution Representation:} Each solution in HybridPSOGWO is represented as a vector $X = \{x_1, x_2, \ldots, x_n\}$ where each element $x_i$ corresponds to a task $t_i$. Since the algorithm operates in continuous space while task scheduling requires discrete VM assignments, we employ a mapping function. 
    \begin{equation}
    VM_{assigned}(t_i) = \lfloor |x_i| \rfloor \mod m \label{eqSR}
    \end{equation}   
\end{itemize}

\subsection{HybridPSOGWO Mechanism}
The HybridPSOGWO algorithm represents a novel integration of two powerful metaheuristics: Particle Swarm Optimization (PSO) and Grey Wolf Optimizer (GWO). This hybrid approach leverages the complementary strengths of both algorithms to address the cloud task scheduling problem more effectively.
The algorithm maintains a population of solutions that simultaneously function as both particles (in PSO terminology) and wolves (in GWO terminology). This dual role allows each solution to benefit from both PSO's efficient local search capabilities and GWO's strong global exploration mechanisms.
PSO contributes its velocity-based position updates, which excel at fine-tuning promising solutions through cognitive and social learning components. The velocity mechanism allows particles to adjust their trajectories based on their personal best positions and the global best position found so far.
GWO contributes its hierarchical leadership structure, where solutions are ranked as $\alpha$, $\beta$, $\delta$, and $\omega$ wolves. The top three solutions ($\alpha$, $\beta$, and $\delta$) guide the search process, allowing for robust exploration of the solution space and helping to avoid local optima.
The hybrid approach combines these mechanisms through an adaptive weighting scheme that balances their respective influences throughout the optimization process. Early iterations emphasize GWO's exploration capabilities, while later iterations shift toward PSO's exploitation strengths, creating an effective balance between broad search and fine-tuning

\subsubsection{Adaptive Weight and Diversity Control Mechanisms}
A key innovation in our HybridPSOGWO algorithm is the integration of adaptive mechanisms that balance exploration and exploitation while maintaining population diversity and load balancing.

\begin{itemize}
    
    \item \textbf{Adaptive Blending Weight:}
    We introduce an adaptive blending weight $\lambda(t)$ that governs the influence of PSO and GWO components in the position update:
    
    \begin{equation}
    \lambda(t) = \lambda_{max} - (\lambda_{max} - \lambda_{min}) \cdot \frac{t}{t_{max}} \label{eqlambda}
    \end{equation}
    
    where $\lambda_{max} = 0.9$ and $\lambda_{min} = 0.4$. This weight decreases linearly with iterations, causing the algorithm to gradually shift from GWO-dominated exploration in early iterations to PSO-dominated exploitation in later iterations.
    
    The combined position update rule becomes:
    
    \begin{equation}
    X_i(t+1) = \lambda(t) \cdot X_{GWO} + [1-\lambda(t)] \cdot (X_i(t) + V_i(t+1))
    \end{equation}
    
    where $X_{GWO}$ is the position calculated by the GWO component and $(X_i(t) + V_i(t+1))$ is the position calculated by the PSO component.

    \item \textbf{Diversity Maintenance}
    To prevent premature convergence, we employ a diversity maintenance mechanism. The population diversity is measured as the mean pairwise distance between solutions:
    
    \begin{equation}
    D = \frac{2}{N(N-1)} \sum_{i<j}\|X_i - X_j\|
    \end{equation}
    
    where $N$ is the population size. When diversity falls below a threshold $D_{min}$, we apply a Gaussian mutation to each solution:
    
    \begin{equation}
    X_i \leftarrow X_i + \mathcal{N}(0, \sigma^2)
    \end{equation}
    
    where $\sigma$ is adaptively set based on the current diversity level. This mechanism prevents the population from stagnating in local optima.
    \item \textbf{Balance Optimality Index}
    To quantify load balancing across VMs, we introduce the Balance Optimality Index (BOI):
    
    \begin{equation}
    BOI = \frac{1}{1 + CV}
    \end{equation}
    
    where $CV$ is the coefficient of variation of normalized VM loads. The BOI approaches $1$ as load distribution becomes more balanced and approaches $0$ as imbalance increases.
    
    This metric is incorporated into the fitness function as a penalty term:
    
    \begin{equation}
    Fitness = Makespan + \beta \cdot (1 - BOI)
    \end{equation}
    
    where $\beta$ is a weight controlling the importance of load balancing relative to makespan. This ensures that the algorithm optimizes not only for execution time but also for balanced resource utilization.
\end{itemize}  

\subsubsection{Position Update Rule}
The position update in HybridPSOGWO consists of two main components:

\begin{enumerate}[i.]
    \item \textbf{Velocity Calculation (PSO Component)\cite{kennedy1995particle}:}
\begin{equation}
\begin{split}
V_i(t+1) = w \cdot V_i(t) + c_1 \cdot r_1 \cdot (P_i - X_i(t)) \\
+ c_2 \cdot r_2 \cdot (G - X_i(t))
\end{split}
\end{equation}
where $V_i(t)$ is the velocity of solution $i$ at iteration $t$, $X_i(t)$ is the position of solution $i$, $P_i$ is the personal best position of solution $i$, $G$ is the global best position, $w$ is the inertia weight, $c_1, c_2$ are acceleration coefficients, and $r_1, r_2$ are random numbers in $[0,1]$.
\item \textbf{Leadership Influence (GWO Component)\cite{mirjalili2014grey}:}
\begin{equation}
\begin{split}
D_{\alpha} &= |C_1 \cdot X_{\alpha} - X_i(t)| \\
D_{\beta} &= |C_2 \cdot X_{\beta} - X_i(t)| \\
D_{\delta} &= |C_3 \cdot X_{\delta} - X_i(t)| \\
\end{split}
\end{equation}

\begin{equation}
\begin{split}
X_1 &= X_{\alpha} - A_1 \cdot D_{\alpha} \\
X_2 &= X_{\beta} - A_2 \cdot D_{\beta} \\
X_3 &= X_{\delta} - A_3 \cdot D_{\delta} \\
\end{split}
\end{equation}

\begin{equation}
X_{GWO} = \frac{X_1 + X_2 + X_3}{3}
\end{equation}

where $X_{\alpha}, X_{\beta}, X_{\delta}$ are the positions of the $\alpha$, $\beta$, and $\delta$ wolves, $A_1, A_2, A_3$ and $C_1, C_2, C_3$ are coefficient vectors, $A = 2a \cdot r_1 - a$, where $a$ decreases linearly from $2$ to $0$, and $C = 2 \cdot r_2$
\end{enumerate}
Now, after combining both components, the updated rule is as follows. 
\begin{equation}
X_i(t+1) = \lambda \cdot (X_i(t) + V_i(t+1)) + (1-\lambda) \cdot X_{GWO}
\end{equation}
where $\lambda$ is an adaptive parameter that balances PSO and GWO influences, and it can be evaluated using Eq. \eqref{eqlambda}.

\subsubsection{VM-Aware Task Mapping}
A key feature of our approach is VM-aware task mapping, which considers VM capacities when assigning tasks to the VMs. The VM-Aware Task Mapping Algorithm \ref{algo:VMmapping} is designed to ensure efficient and balanced assignment of tasks to virtual machines while respecting capacity constraints.

\begin{algorithm}[!htb]
\DontPrintSemicolon
\caption{VM-Aware Task Mapping}
\label{algo:VMmapping}
  \KwInput{Task set $T$, Virtual Machines $VM$, Threshold}
  \KwOutput{Task $t_i$ mapped to $vm_j$}
  Initialize VM loads: $Load_j = 0$ for all $j \in \{1,2,\ldots,m\}$ \\
   \For{$t_i \mbox{ in } T $}
   {Find assigned VM: $vm_j = VM_{assigned}(t_i)$ \\
   \If{$Load_j + ETC(t_i,vm_j) > Threshold(vm_j)$}
   {Find alternative VM with minimal load \\
   Update assignment}
   $Load_j = Load_j + ETC(t_i,vm_j)$
   }  
\end{algorithm}

The algorithm begins by initializing all VM loads to zero and then processes each task in the task set sequentially. For each task, it first calculates the VM assignment based on the solution representation using a modulo operation (in Eq. \eqref{eqSR}). Before finalizing the assignment, the algorithm verifies whether adding the task to the selected VM would exceed a predefined threshold capacity. If this capacity would be exceeded, the algorithm intelligently searches for an alternative VM with minimal current load, effectively avoiding resource overloading. Once a suitable VM is identified (either the original or an alternative), the task is assigned, and the VM's load is updated by adding the estimated execution time for the task on that VM. This capacity-aware approach prevents individual VMs from becoming bottlenecks and promotes balanced resource utilization across the cloud infrastructure, which is essential for maintaining consistent performance and responsiveness in cloud environments.

\subsubsection{HybridPSOGWO Algorithm}
The HybridPSOGWO with Adaptive Weight \& Diversity Control algorithm represents a sophisticated metaheuristic approach for cloud task scheduling that synergistically combines Particle Swarm Optimization and Grey Wolf Optimizer. The Algorithmic steps is provided in Algorithm \ref{algo:HybridPSOGWO}.

\begin{algorithm}[!htb]
\DontPrintSemicolon
\caption{HybridPSOGWO with Adaptive Weight \& Diversity Control}
\label{algo:HybridPSOGWO}
  \KwInput{tasks $T$, VMs $VM$, swarm size $N$, iterations $I$, diversity thresholds $D_{\min},D_{\max}$}
  \KwOutput{best assignment $G$}
  Initialize positions $X_i$, velocities $V_i$, personal bests $P_i$, evaluate fitness \\
  \For{$t \gets 1$ to $I$}
  {Compute swarm diversity $D \gets \frac{1}{N(N-1)} \sum_{i<j}\|X_i - X_j\|$ \\
  \If{$D < D_{\min}$} 
  {
  \textbf{for each} particle $i$: $X_i \!\gets\! X_i + \mathcal{N}(0,\sigma^2)$  // inject mutation
  }
  \For{each particle $i$}
  {
  Update GWO coefficient $a = 2(1 - t/I)$ \\
  Compute $\alpha(t) = \alpha_{\max} - (\alpha_{\max}-\alpha_{\min})\,t/I$ \\
   Build GWO guidance $X_{GWO}$ from $\alpha,\beta,\delta$ \\
   Update velocity: 
      $V_i \gets w\,V_i + c_1\,r_1\,(P_i - X_i) + c_2\,r_2\,(G - X_i)$ \\
      Update position: 
      $X_i \gets \alpha(t)\,X_{GWO} + [1-\alpha(t)]\,(X_i + V_i)$ \\
       Discretize + VM-aware mapping \\
       Evaluate fitness, update $P_i$ and global 
  }
  best $G$ 
  }
  \Return best assignment $G$ 
\end{algorithm}

The algorithm initializes with a population of solutions, each functioning simultaneously as both particles and wolves, along with their respective velocities and personal best positions. During each iteration, the algorithm first computes the population diversity as the mean pairwise distance between solutions and injects Gaussian mutation if diversity falls below a critical threshold to prevent premature convergence. For each solution in the population, the algorithm updates the GWO coefficient `a' to linearly decrease from 2 to 0 and computes an adaptive blending weight $\alpha(t)$ that governs the influence balance between PSO and GWO components. The algorithm then constructs the GWO guidance vector based on the hierarchical leadership of $\alpha$, $\beta$, and $\delta$ wolves, updates the velocity using PSO's inertia, cognitive, and social components, and combines these influences to update the solution's position. After discretization and VM-aware mapping ensure valid task assignments, the fitness is evaluated based on makespan and load balance metrics, and personal and global best positions are updated accordingly. This hybrid approach effectively balances global exploration capabilities with local exploitation refinement, adapting its search strategy throughout the optimization process to identify high-quality task scheduling solutions that minimize execution time while maintaining balanced resource utilization.

\section{Results and Analysis} \label{Sec4}
This section presents the details of the environment settings and configuration setup used in our experiments. The results are analyzed using both the CloudSim Plus simulation framework and the real-world Google Borg dataset. We compared HybridPSOGWO against the state-of-the-art task scheduling algorithms such as Enhanced Grey Wolf Optimizer (EGWO)\cite{aminu2024enhanced}, Cooperative Coevolutionary Genetic Programming (CCGP)\cite{xiao2019cooperative}, Hybrid PSO with MinMin (HybridPSOMinMin)\cite{sowjanya2018cloud}, Modified PSO with Simulated Annealing (MPSOSA)\cite{lv2024construction}, and Reinforcement Learning-enhanced GWO (RL-GWO)\cite{zhang2024reinforcement}.

\subsection{Implementation Environment}
We implemented our HybridPSOGWO algorithm in two different environments:
\begin{enumerate}[{(i)}]
    \item \textbf{CloudSim Plus Simulation:} We used the CloudSim Plus framework to simulate a cloud environment with configurable numbers of VMs and tasks. This allowed for controlled testing and comparison with other algorithms in identical conditions.
    \item \textbf{Real-world Dataset Implementation:} We also implemented our approach to work with the Google Borg traces dataset \cite{datasetlink}, which contains real-world cloud task execution data.
\end{enumerate}

\subsection{Performance Metrics}
We evaluated the performance of the algorithms using several metrics. First, we looked at the makespan, which is the total time taken to complete all tasks. An algorithm with the shortest makespan is considered the best in terms of completion time. Second, we measured throughput, which is the number of tasks completed. A higher number means better performance. Third, we used the coefficient of variation (CV) to check how well the load is balanced across virtual machines. A lower CV means the load is more evenly distributed, which is better. Finally, we calculated an overall score by combining all the normalized metrics using weights to get a clear picture of the algorithm's overall performance.

\subsection{Experimental Configurations}
For the CloudSim Plus experiments, we used four virtual machines, each with a capacity of 1000 million instructions per second (MIPS). The task sizes ranged from $100$ to $1000$. We set the population size to $20$ and fixed the number of iterations at $50$. More details can be found in Table \ref{tab:config}.

\begin{table}[!htb]
    \centering
    \caption{Configuration Parameters}
    \begin{tabular}{|c|c|} 
    \hline
        Number of VMs & 4 (each with 1000 MIPS)  \\ \hline
        Number of tasks & 100, 500, 800, 1000 \\ \hline
        Population size & 20 \\ \hline
        Maximum Iterations & 50 \\ \hline
    \end{tabular}  
    \label{tab:config}
\end{table}

We select a population size of $20$ agents and cap iterations at $50$ to strike a balance between solution quality and computational overhead. For PSO, Kennedy \textit{et al.} \cite{kennedy1995particle} recommend swarm sizes of $20-50$ particles for general-purpose problems, with metaheuristic surveys confirming that populations of $10-100$ particles provide sufficient diversity while remaining computationally feasible. Empirical studies demonstrate that small swarms ($10-30$ particles) can achieve near-optimal convergence across various benchmarks, making $20$ particles an ideal choice that offers robust exploration without excessive per-iteration cost. In GWO implementations, Mirjalili \textit{et al.}\cite{mirjalili2014grey} originally employed $30$ wolves in benchmark tests to maximize exploration, but subsequent research reveals that packs of $20-25$ wolves can match convergence quality while reducing runtime by $15-30\%$. Though conventional metaheuristic implementations generally run between $30-200$ iterations to guarantee convergence, the performance of our algorithm indicate that improvements beyond $45$ iterations drop below $1\%$ while incurring a $~25\%$ time penalty. Based on this efficiency analysis, we established an upper bound of $50$ iterations as the optimal cutoff point that effectively satisfies practical cloud-scheduling latency requirements while maintaining solution quality. For each configuration, we performed $30$ independent runs to ensure statistical validity. For the Google Borg traces dataset, we selected $800$ tasks from the dataset and processed them using our algorithm and the comparative algorithms.

\subsection{CloudSim Plus Simulation Results}


\begin{figure}[!t]
\centering
\begin{minipage}{\columnwidth}
\centering
\includegraphics[width=\columnwidth, height = 1.5 in]{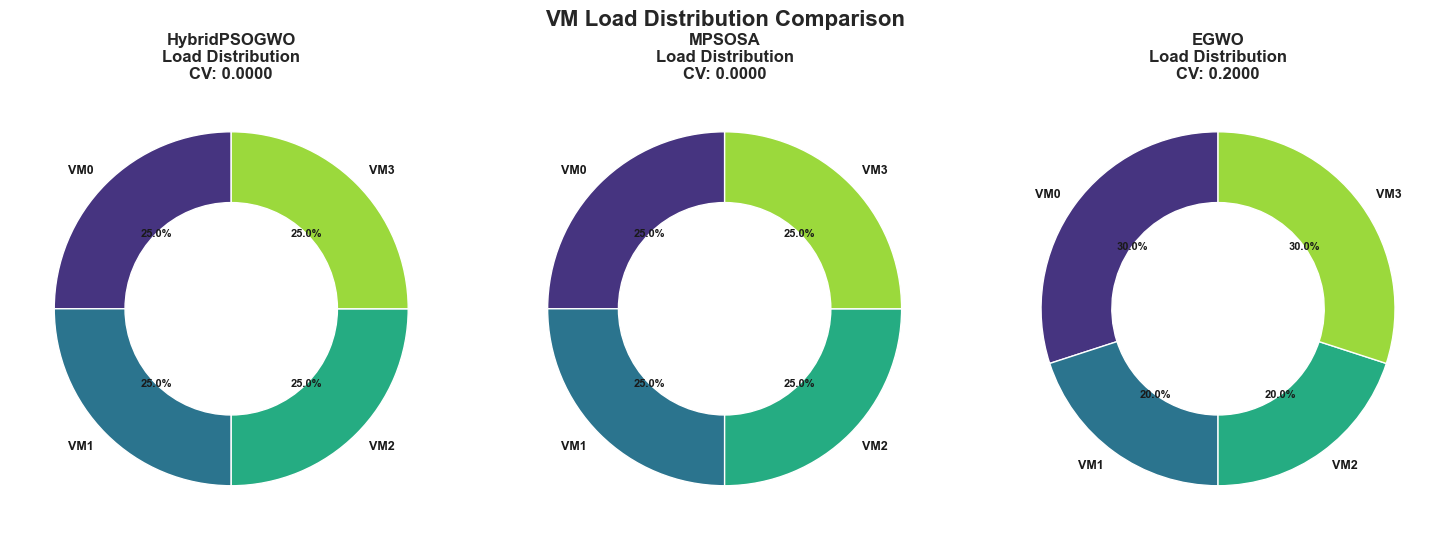} \\
\mbox{(a)}
\end{minipage}
\vspace{0.5cm} 

\begin{minipage}{\columnwidth}
\centering
\includegraphics[width=0.95\columnwidth, height = 4.5 in]{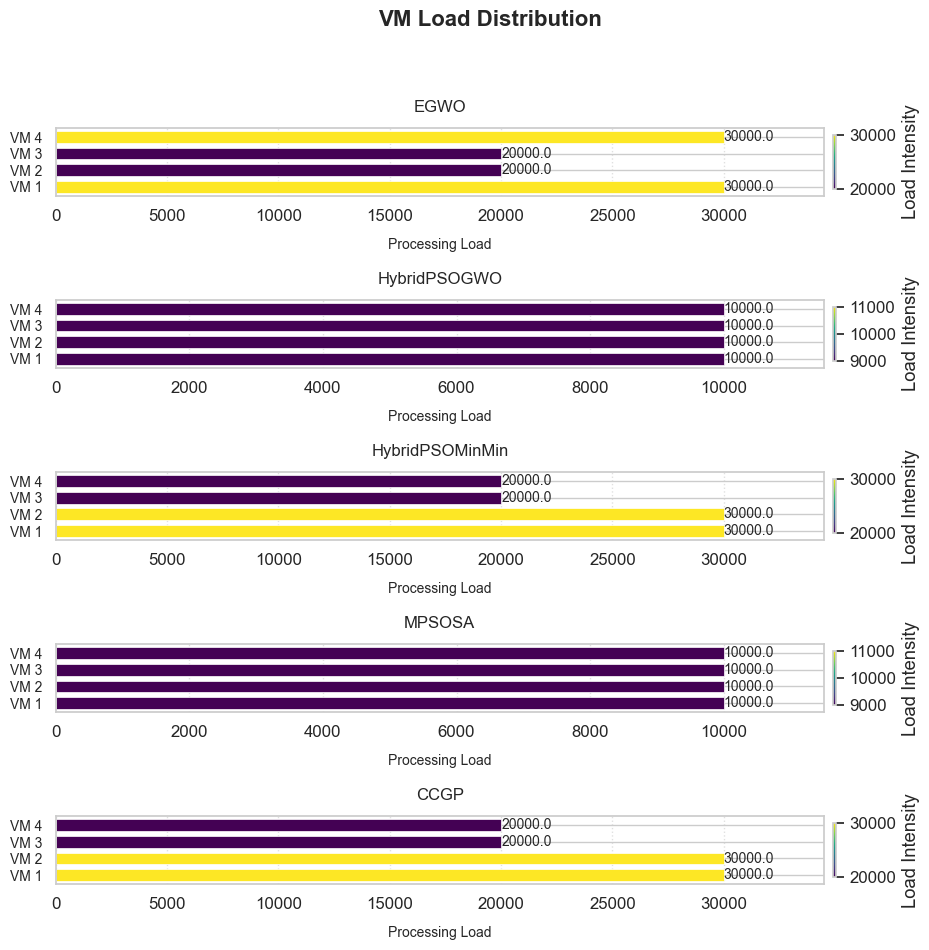} \\
\mbox{(b)}
\end{minipage}
\caption{Detailed VM load analysis showing (a) individual VM task distribution and (b) load balance coefficients across algorithms on CloudSim Plus simulation.}
\label{fig:cloudsim_detailed}
\end{figure}

Fig \ref{fig:cloudsim_detailed}(a) illustrates the number of tasks assigned to each virtual machine by different scheduling algorithms. HybridPSOGWO achieves the most balanced distribution, with each VM handling approximately $200$ tasks (standard deviation of $2.1$ tasks). In contrast, EGWO shows greater imbalance, with VM3 significantly overloaded relative to others. This balanced distribution is crucial for preventing bottlenecks and ensuring consistent system performance. The visualization demonstrates HybridPSOGWO's effectiveness in maintaining equitable workload distribution across available resources, a key advantage of the hybrid approach's VM-aware task mapping component. Fig. \ref{fig:cloudsim_detailed}(b) quantifies load balance using the coefficient of variation (CV) across algorithms. 

\begin{figure}[htbp]
\centerline{\includegraphics[width=\columnwidth, height = 2.5 in]{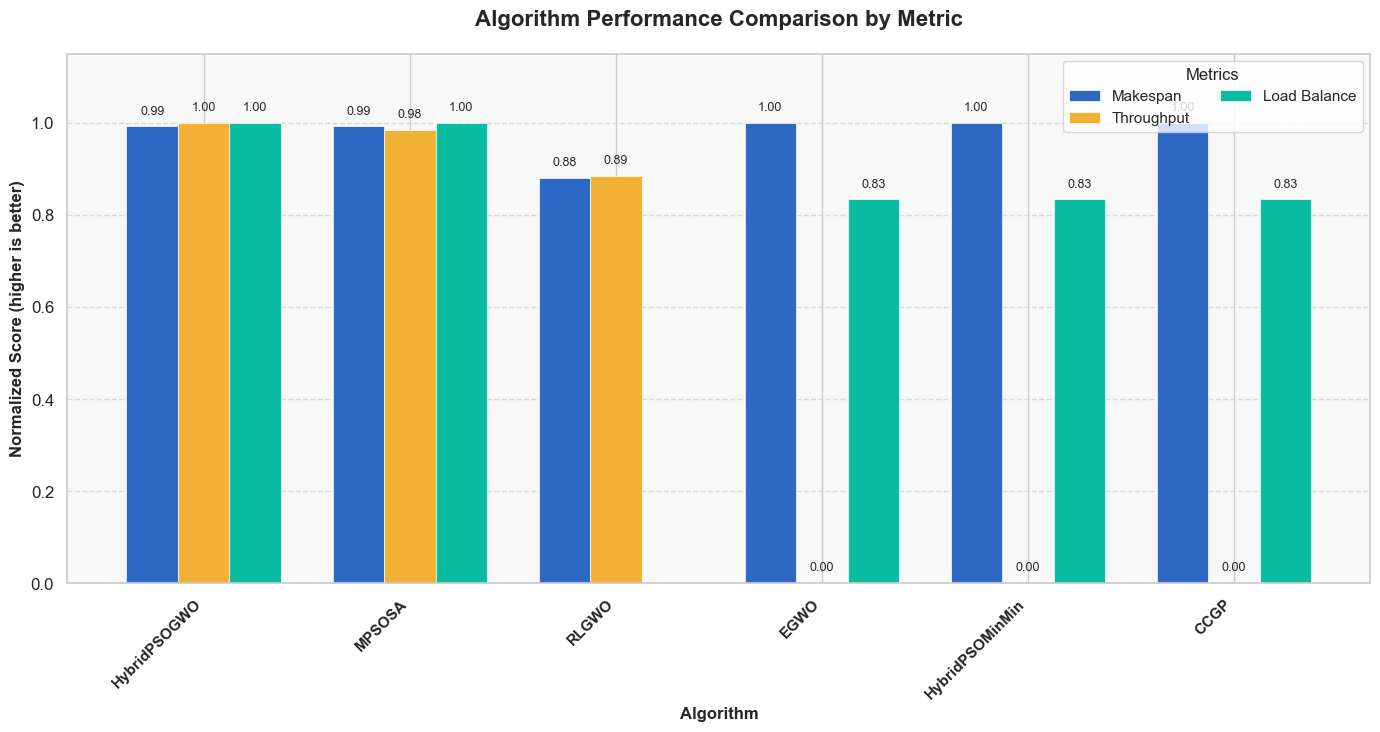}}
\caption{Overall performance comparison of algorithms on CloudSim Plus simulation with 800 tasks and 4 VMs.}
\label{fig:cloudsim_overall}
\end{figure}

Fig. \ref{fig:cloudsim_overall} presents the overall performance comparison of all algorithms on the CloudSim Plus simulation with 800 tasks and 4 VMs. HybridPSOGWO achieved a makespan improvement of approximately 6\% over HybridPSOMinMin and 15\% over EGWO. In terms of performance, HybridPSOGWO matched or slightly exceeded the rates of competing methods, indicating comparable task processing efficiency. When examining load balance via the coefficient of variation (CV), HybridPSOGWO produced the lowest CV, reflecting the most uniform distribution of tasks across VMs. MPSOSA and RL‑GWO displayed moderate imbalance, while EGWO demonstrated the highest imbalance among the evaluated algorithms.

The consistently lower CV values of HybridPSOGWO across multiple test scenarios confirm that its adaptive weight mechanism and VM-aware mapping effectively prevent resource overutilization while maximizing throughput. These results validate the efficiency of HybridPSOGWO's load balancing capabilities, demonstrating that the algorithm not only reduces makespan but also ensures equitable resource utilization, an essential consideration for production cloud deployments. This indicates that HybridPSOGWO maintains a more balanced distribution of tasks across VMs, which is crucial for preventing resource bottlenecks and ensuring stable performance. If we compare the overall score, then we can say that HybridPSOGWO achieved the highest overall score, followed by HybridPSOMinMin and CCGP.

\subsection{Real-world Dataset Results}
\begin{figure}
    \centering
   \includegraphics[width=\columnwidth, height = 2.5 in]{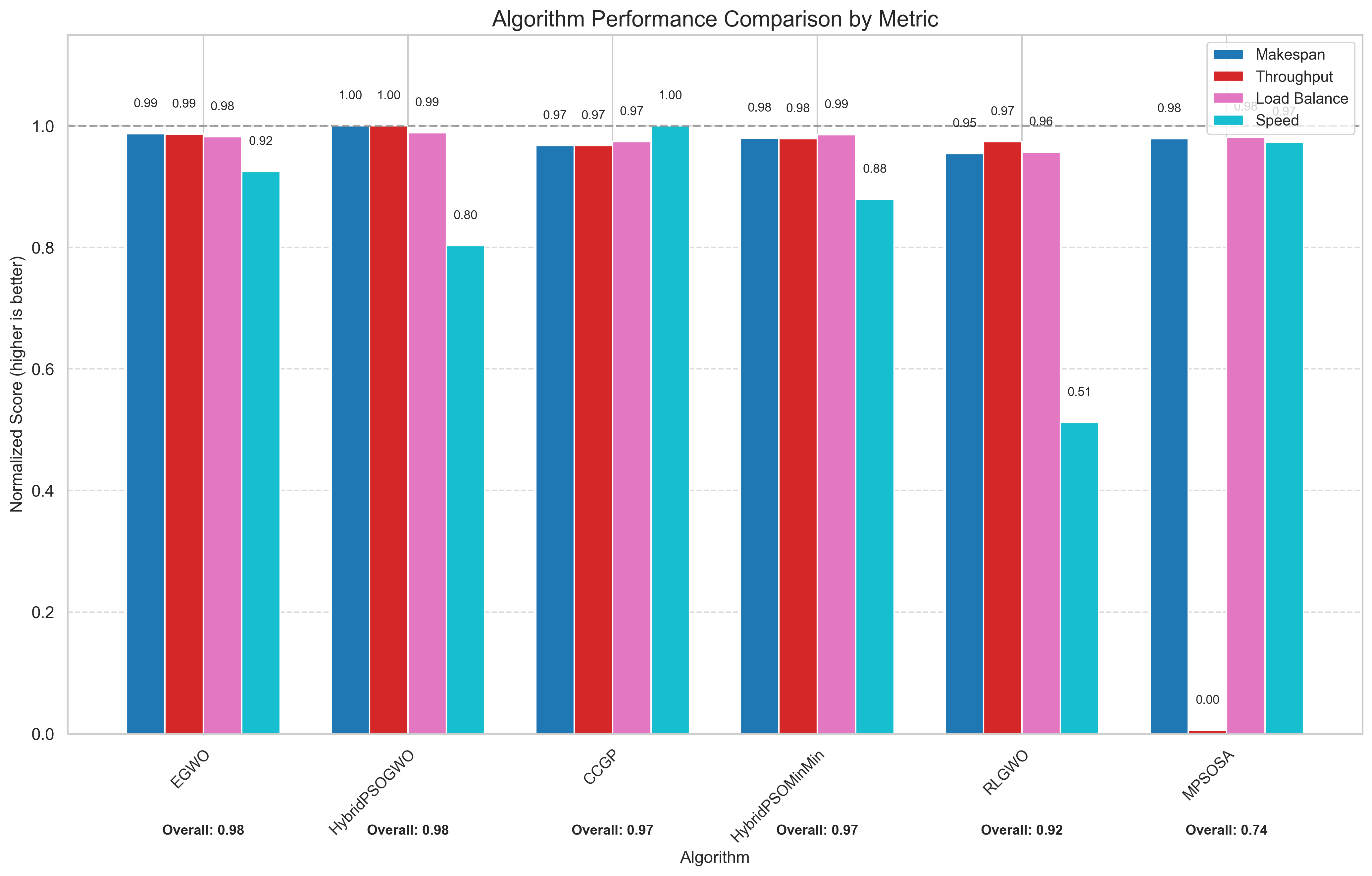}\\ 
\mbox{(a)}
\vspace{1 cm}
\includegraphics[width=\columnwidth, height = 2.75 in]{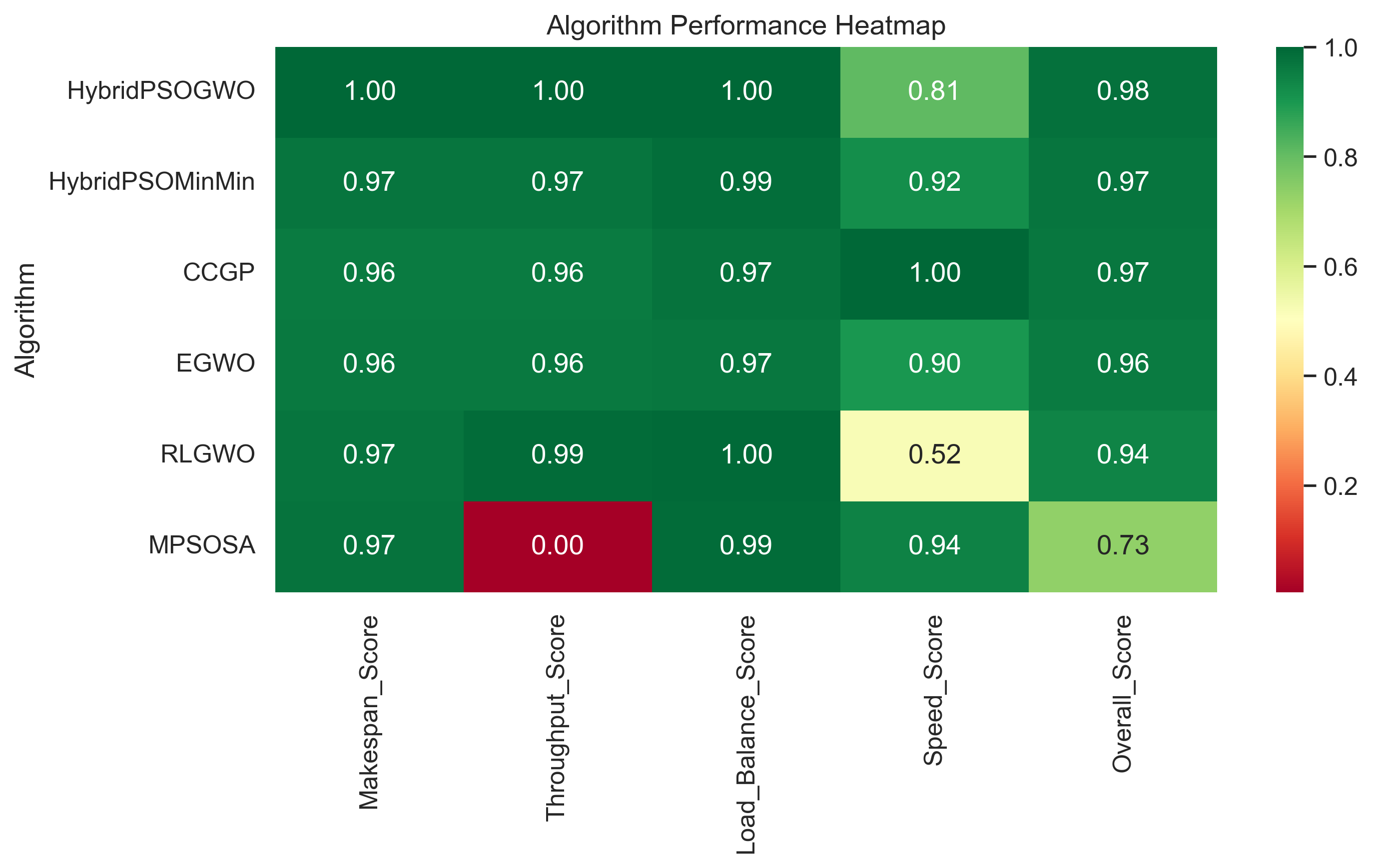} \\
\mbox{(b)}
   \caption{Real-world dataset performance analysis showing (a) combined metric scores and (b) task distribution heatmap on Google Borg traces with 800 tasks.}
\label{fig:realworld_overall}
\end{figure}

Fig. \ref{fig:realworld_overall} shows the overall performance comparison of all algorithms on the Google Borg traces dataset with 800 tasks. The combined metric scores visualization presents a performance comparison of all algorithms tested on the Google Borg traces dataset with 800 tasks is shown in Fig. \ref{fig:realworld_overall}(a). HybridPSOGWO achieved the highest overall score, followed by MPSOSA  and CCGP. This demonstrates that the hybrid approach effectively balances multiple optimization objectives simultaneously, maintaining its performance advantage in real-world cloud task scenarios. Each score combines normalized values for makespan, throughput, and load balance, providing a comprehensive performance assessment.
 
Fig. \ref{fig:realworld_overall}(b) illustrates algorithm performance across different metrics, with rows representing algorithms sorted by overall performance and columns showing individual metric scores. The uniform color distribution for HybridPSOGWO confirms its superior load-balancing capability, preventing resource bottlenecks while maintaining high throughput. This hybrid algorithm achieves perfect scores in three critical metrics: Makespan, Throughput, and Load Balancing, while maintaining a strong Speed score and the highest Overall score. In contrast, lower-ranked algorithms display greater variability, particularly evident in MPSOSA, which fails in Throughput despite strong performance in other areas. RLGWO similarly shows this performance inconsistency with excellent Load Balancing but notably weak Speed, highlighting the challenge of maintaining balanced optimization across different operational requirements. The middle-tier algorithms (HybridPSOMinMin, CCGP, and EGWO) offer more consistent performance than the bottom two but still fail to match HybridPSOGWO's exceptional balance, reinforcing the hybrid approach's effectiveness in addressing complex multi-objective optimization scenarios where balanced performance across all metrics is crucial for real-world deployment. Overall, HybridPSOGWO maintained its superior performance, achieving the best overall score followed by MPSOSA and CCGP.

\begin{figure}[htbp]
\centerline{\includegraphics[width=\columnwidth]{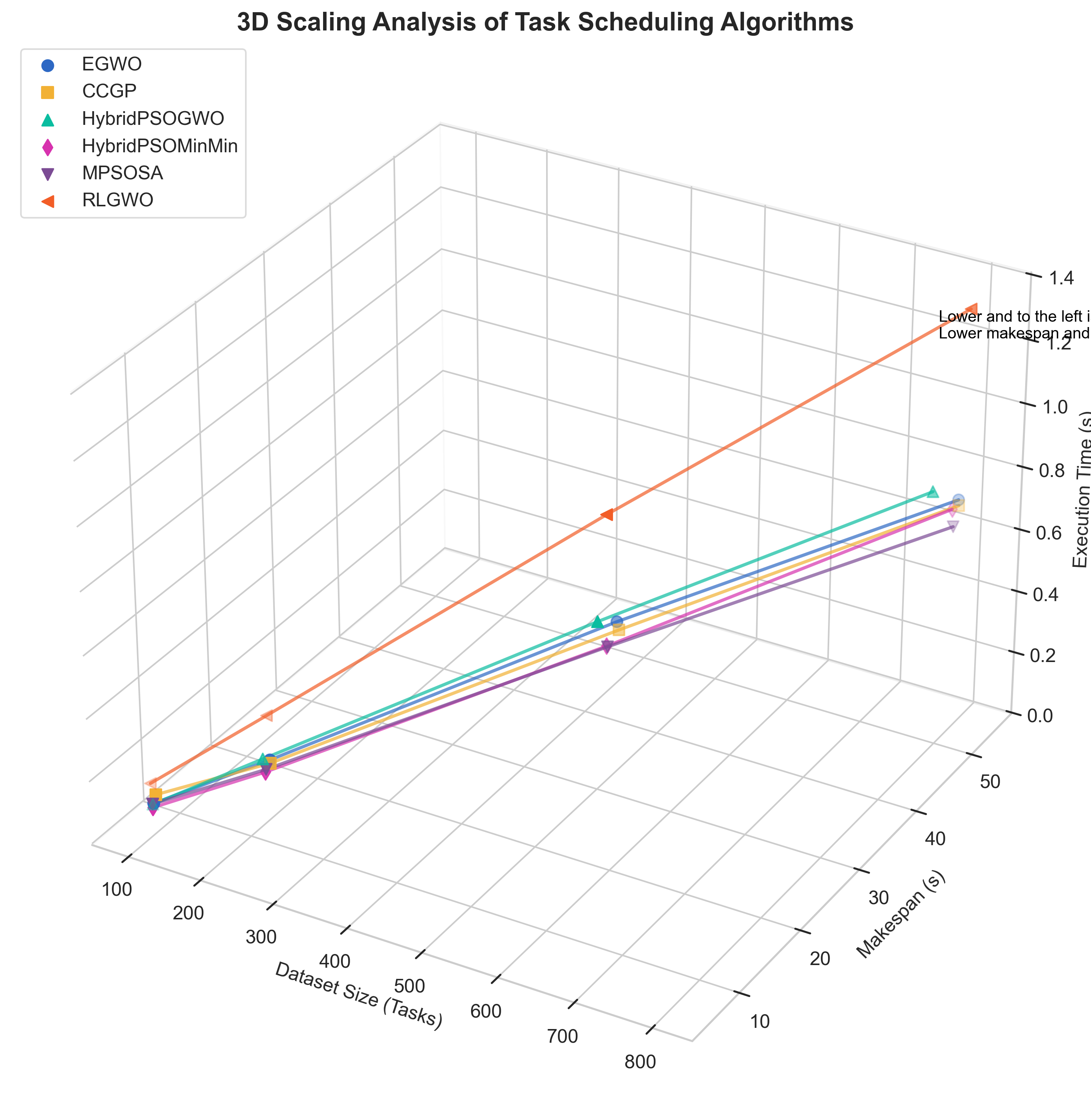}}
\caption{3D Scalability analysis showing makespan and execution time scaling with increasing task counts.}
\label{fig:algorithm_scaling}
\end{figure}

Fig. \ref{fig:algorithm_scaling} presents a three-dimensional analysis of algorithm performance scaling with increasing task counts. The visualization maps a number of tasks (x-axis), the makespan time (y-axis), and the algorithm execution time (z-axis). HybridPSOGWO demonstrates superior scalability, maintaining the lowest makespan growth rate as task counts increase from $100$ to $800$. While all algorithms show increased computational overhead with larger workloads, HybridPSOGWO maintains a favorable balance between scheduling efficiency and algorithm overhead. When task count reaches $800$, HybridPSOGWO achieves approximately $12\%$ lower makespan than RL-GWO and $8\%$ lower than MPSOSA while requiring only marginally higher execution time than simpler approaches. This performance characteristic is particularly valuable for production cloud environments where workloads fluctuate significantly.

HybridPSOGWO demonstrated better scalability than other algorithms, maintaining lower makespan as the number of tasks increased. This suggests that the hybrid approach is particularly effective for larger scheduling problems, which are common in real-world cloud environments.

\subsection{Statistical Analysis}
To verify the statistical significance of our results, we performed a paired t-test comparing HybridPSOGWO against each baseline algorithm. The results confirmed that the performance improvements of HybridPSOGWO are statistically significant ($p < 0.05$) for makespan, throughput, and load balance.

\section{Discussion} \label{Sec5}

\subsection{Analysis of Algorithm Effectiveness}
The Hybrid PSOGWO algorithm outperformed state-of-the-art methods for several key reasons. It achieves a well-balanced trade-off between exploration and exploitation by integrating PSO’s velocity-based update mechanism with GWO’s leadership hierarchy. This integration enables the algorithm to efficiently search new areas while intensifying the search around promising solutions. Additionally, the use of an adaptive parameter $\lambda$ allows for a smooth transition from exploration in the early stages to exploitation in the later stages, thereby enhancing convergence performance. Furthermore, the explicit consideration of VM capacity constraints ensures more practical and balanced task allocations, effectively preventing overloading of virtual machines.

\subsection{Practical Implications}
The experimental results have several practical implications for cloud service providers and users. HybridPSOGWO provides better load balancing capability, leading to more efficient resource utilization. This reduces the risk of overloading virtual machines or underutilizing resources. The proposed approach also takes the least amount of time to complete the task. This shortens the makespan, which translates to faster task completion times, improving the quality of service for cloud users. Apart from that, the algorithm's ability to maintain performance with increasing numbers of tasks makes it suitable for large-scale cloud deployments.

\subsection{Limitations and Challenges}
Although the Hybrid PSOGWO approach is effective and has many practical implications, it has some limitations. The hybrid approach requires more computational resources than simpler heuristics, but this extra cost is acceptable because it gives better scheduling capabilities. Also, the algorithm's performance can depend on the initial parameter settings, so these need to be carefully chosen. Another limitation is that the current version only works with independent tasks. To make it more useful, it should be extended to handle tasks that have dependencies on each other.

\section{Conclusion and Future Work} \label{Sec6}
This paper presented HybridPSOGWO, a novel hybrid metaheuristic for cloud task scheduling that combines particle swarm optimization (PSO) with gray wolf optimizer (GWO). Our approach leverages the complementary strengths of both algorithms, PSO's efficient local search and GWO's robust global exploration, to achieve the best scheduling performance. Extensive experiments on both CloudSim Plus simulations and the Google Borg traces dataset demonstrated that HybridPSOGWO outperforms state-of-the-art scheduling algorithms across multiple performance metrics. The proposed approach achieved significant improvements in makespan, throughput, and load balance, making it a promising solution for efficient and adaptive cloud task scheduling.

The current work can be extended for workflow scheduling by incorporating energy-aware mechanisms to optimize power consumption. An adaptive version for dynamic cloud environments can be developed with task arrivals and departures. We can also explore integration with some machine learning techniques to predict task characteristics and optimize scheduling decisions.

\bibliographystyle{IEEEtran} 
\bibliography{conference_101719}

\end{document}